\documentclass{aastex61}
\usepackage{lineno}
\usepackage{xspace}
\usepackage{soul}
\usepackage{color}
\usepackage{csquotes}
\newcommand\aastex{AAS\TeX}

\usepackage{xcolor}
\definecolor{burntorange}{rgb}{0.8, 0.33, 0.0}

\newcommand{\CSPAR}{\affiliation{Center for Space Plasma and Aeronomic Research, University of Alabama in Huntsville, Huntsville, AL 35899, USA}}
\newcommand{\LSU}{\affiliation{Department of Physics and Astronomy, Louisiana State University, Baton Rouge, LA 70803 USA}}
\newcommand{\MSFCAstro}{\affiliation{ST12 Astrophysics Branch, NASA Marshall Space Flight Center, Huntsville, AL 35812, USA}}
\newcommand{\SPA}{\affiliation{Department of Space Science, University of Alabama in Huntsville, 320 Sparkman Drive, Huntsville, AL 35899, USA}}
\newcommand{\USRA}{\affiliation{Science and Technology Institute, Universities Space Research Association, Huntsville, AL 35805, USA}}

\newcommand{\Ioffe}
{\affiliation{Ioffe Institute, 26 Politekhnicheskaya, St. Petersburg, 194021, Russia}}
\newcommand{\Cagliari}
{\affiliation{
Dipartimento di Fisica, Universit\`{a} degli Studi di Cagliari, SP Monserrato-Sestu, km 0.7, I-09042 Monserrato, Italy}}

\shorttitle{\aastex\ The BOAT}

\begin{document}
\AuthorCallLimit=999
\title{GRB\,221009A\\The BOAT}

%direct contributing individuals
\author[0000-0002-2942-3379]{Eric~Burns}\thanks{Corresponding author: ericburns@lsu.edu}
\LSU

\author[0000-0002-2208-2196]{Dmitry~Svinkin}
\Ioffe

\author{Edward~Fenimore}
\affiliation{Los Alamos National Laboratory, P.O. Box 1663, Los Alamos, NM, 87545}

\author[0000-0003-2902-3583]{D.~Alexander~Kann}
\affiliation{Hessian Research Cluster ELEMENTS, Giersch Science Center, Max-von-Laue-Stra{\ss}e 12, Goethe University Frankfurt, Campus Riedberg, 60438 Frankfurt am Main, Germany}

\author[0000-0001-6991-7616]{Jos\'e~Feliciano~Ag\"u\'i~Fern\'andez}
\affiliation{Instituto de Astrof\'isica de Andaluc\'ia (IAA-CSIC), Glorieta de la Astronom\'ia s/n, 18008 Granada, Spain}

\author[0000-0002-1153-6340]{Dmitry~Frederiks}
\Ioffe

\author[0000-0003-0761-6388] {Rachel~Hamburg}
\affiliation{Universit\'e Paris-Saclay, CNRS/IN2P3, IJCLab, 91405 Orsay, France}

\author[0000-0001-8058-9684]{Stephen~Lesage}
\SPA
\CSPAR

\author[0000-0003-4813-8378]{Yuri~Temiraev}
\Ioffe

\author[0000-0003-0292-6221]{Anastasia~Tsvetkova}
\Ioffe
\Cagliari

%GBM, Konus teams, alphabetical
\author[0000-0001-9935-8106]{Elisabetta~Bissaldi}
\affiliation{Dipartimento Interateneo di Fisica, Politecnico di Bari, Bari, Italy}
\affiliation{Istituto Nazionale di Fisica Nucleare, Sezione di Bari, Bari, Italy}

\author[0000-0003-2105-7711]{Michael~S.~Briggs}
\SPA
\CSPAR

\author[0000-0003-1835-570X]{Sarah~Dalessi}
\SPA
\CSPAR

\author[0000-0003-3248-5447]{Rachel~Dunwoody}
\affiliation{School of Physics, Centre for Space Research, Science Center North, University College Dublin, Dublin 4, Ireland}

\author[0000-0002-0186-3313]{Cori~Fletcher}
\USRA

\author[0000-0002-0587-7042]{Adam~Goldstein}
\USRA

\author[0000-0002-0468-6025]{C.~Michelle~Hui}
\MSFCAstro

\author[0000-0001-9556-7576]{Boyan~A.~Hristov}
\CSPAR

\author[0000-0001-9201-4706]{Daniel~Kocevski}
\MSFCAstro

\author[0000-0002-3942-8341]{Alexandra~L.~Lysenko}
\Ioffe

\author[0000-0002-2531-3703]{Bagrat~Mailyan}
\affiliation{Department of Aerospace, Physics and Space Sciences, Florida Institute of Technology, Melbourne, FL 32901, USA}

\author[0000-0002-6849-5009]{Joseph~Mangan}
\affiliation{Laboratoire de Physique des 2 infinis Irène Joliot-Curie – IJCLab, CNRS / Université Paris-Saclay / Université Paris Cité, Bâtiment 104 Rue Henri Becquerel,  91405 Orsay Campus, France}

\author[0000-0002-1477-618X]{Sheila~McBreen}
\affiliation{School of Physics, Centre for Space Research, Science Center North, University College Dublin, Dublin 4, Ireland}

\author[0000-0002-4744-9898]{Judith~Racusin}
\affiliation{NASA Goddard Space Flight Center, Greenbelt, MD 20771, USA}

\author[0000-0001-9477-5437]{Anna~Ridnaia}
\Ioffe

\author[0000-0002-7150-9061]{Oliver~J.~Roberts}
\affiliation{Science and Technology Institute, Universities Space and Research Association, 320 Sparkman Drive, Huntsville, AL 35805, USA.}

\author[0000-0002-0076-5228]{Mikhail~Ulanov}
\Ioffe

\author[0000-0002-2149-9846]{Peter~Veres}
\SPA
\CSPAR

\author[0000-0002-8585-0084]{Colleen~A.~Wilson-Hodge}
\MSFCAstro

\author[0000-0001-9012-2463]{Joshua~Wood}
\MSFCAstro

\begin{abstract}
GRB\,221009A has been referred to as the Brightest Of All Time (the \textit{BOAT}). We investigate the veracity of this statement by comparing it with a half century of prompt gamma-ray burst observations. This burst is the brightest ever detected by the measures of peak flux and fluence. Unexpectedly, GRB\,221009A has the highest isotropic-equivalent total energy ever identified, while the peak luminosity is at the $\sim99$th percentile of the known distribution. We explore how such a burst can be powered and discuss potential implications for ultra-long and high-redshift gamma-ray bursts. By geometric extrapolation of the total fluence and peak flux distributions GRB\,221009A appears to be a once in 10,000\,year event. Thus, it is almost certainly not the BOAT over all of cosmic history, it may be the brightest gamma-ray burst since human civilization began.
\end{abstract}
\keywords{gamma rays:  general, methods: observation}

\section{Introduction} \label{sec:intro} 
Cosmological gamma-ray bursts (GRBs) are the most luminous electromagnetic events identified in the Universe since the Big Bang. GRBs were accidentally discovered in 1967 by the \textit{Vela} series of satellites launched to monitor Earth for atmospheric nuclear detonation signatures following the Partial Nuclear Test Ban Treaty \citep{GRBs_astrophysical_origin_1973}. Through \textit{Vela}, its successors, and instruments designed for astrophysics and planetary research, humanity has monitored the full gamma-ray sky for 55\,years.

Cosmological GRBs arise from bi-polar, relativistic jets powered by compact central engines \citep{zhang2018physics}. These jets undergo internal dissipation releasing the prompt GRB emission in the keV and MeV regimes and subsequently interact with the circumburst material to develop an external shock which releases synchrotron emission across the electromagnetic spectrum, referred to as afterglow. Cosmological GRBs are separated into two overlapping classes based on prompt duration, generally separated by a threshold value of 2\,s \citep{1981Ap&SS..80....3M,Dezalay1991,Kouveliotou1993}, which are now known to have different progenitor systems. Short GRBs arise from neutron star mergers \citep{Goldstein2017,Savchenko2017,GW170817-GRB170817A}; long GRBs arise from collapsars, a rare, fast-rotating subset of core-collapse supernovae \citep{galama1998unusual,cano2017observer}\footnote{Notably, some GRBs have contradicted the 2\,s
divide. Recent exemplars include the temporally short supernova-associated GRB\,200826A \citep{Ahumada2021NatAst,Zhang2021NatAst,Rossi2022ApJ} and
the temporally long, nearby kilonova-associated (i.e. strongly suggestive of a neutron star merger origin) GRB\,211211A
\citep{Rastinejad2022Nature,Troja2022Nature,Gompertz2023NatAst}.}. A small number of detected short GRBs are magnetar giant flares \citep{1979Natur.282..587M,2008ApJ...680..545M,svinkin2021bright,burns2021identification} which are not of interest here as they have a distinct physical origin.

The brightness of prompt GRBs can be quantified through different parameters. The time-integrated brightness at Earth is the fluence, which corresponds to the intrinsic brightness measure E$_{\rm iso}$, the total isotropic-equivalent energetics calculated by assuming an equal fluence over a sphere centered on the source with a radius equal to the luminosity distance from source to Earth \citep{piran1999gamma}. The peak flux corresponds to the highest time-resolved flux in a specified interval of time as measured at Earth, with L$_{\rm iso}$ being the isotropic-equivalent measure of the maximum power output in a specified interval. When the opening angle of the jetted outflow is known, the isotropic-equivalent energetics can be converted to the more accurate collimation-corrected energetics \citep{1999ApJ...519L..17S}. %\textcolor{red}{Citations here?}

The exceptionally bright long GRB\,221009A was discovered by a fleet of satellites on October 9th, 2022 \citep[e.g.][as well as particle detectors on-board \textit{MAVEN}, \textit{GAIA}, \textit{STEREO} (R. Leske, private communication), \textit{ACE} (R. Leske, private communication), and \textit{Voyager 1} (A. Cummings, private communication), though the detection by \textit{Voyager 1} occurred on the 8th]{grb221009a_lesage_fermi_gbm_2023,williams2023grb,grb221009a_konus_2023,Ripa2023GRBAlpha,2022GCN.32660....1G,2022GCN.32661....1X,2022GCN.32805....1K,2022GCN.32751....1L,2022GCN.32657....1P,2022GCN.32650....1U,2022GCN.32663....1L,grb221009_gecam_2023}. The GRB has been observed across the electromagnetic spectrum, from beyond 10\,TeV by LHAASO \citep{2022GCN.32677....1H} down to radio \citep[e.g.][]{williams2023grb,grb221009a_fulton_optical_2023,2023arXiv230204388L,kann2023grb,Malesani2023redshift,o2023structured,t2023}. Other key observations include the first observations of the prompt and afterglow polarization from the same burst \citep{negro2023ixpe}, the (somewhat) surprising \citep{murase2022neutrinos} lack of neutrinos \citep{grb221009a_icecube_collaboration_2023}, and the first disruptive target of opportunity from an out-of-cycle proposal of \textit{JWST} \citep[][also the first JWST observation of a GRB afterglow ever]{grb221009a_jwst_hst_2023}.

GRB\,221009A was initially flagged as having the highest prompt fluence and peak flux ever identified by both the \textit{Fermi} Gamma-ray Burst Monitor (GBM) and the Konus-\textit{Wind} (Konus) instruments \citep{2022GCN.32636....1V,2022GCN.32668....1F}. As each instrument has individually identified $\sim$3,500\,GRBs, comprising the two largest prompt GRB samples, GRB\,221009A was referred to as the Brightest Of All Time, \textit{The BOAT}. Given the age and size of the Universe it is exceedingly unlikely that GRB\,221009A is truly the brightest ever. Colloquially, this statement refers to the brightest prompt phase from identified GRBs. 

This paper quantifies the validity of the BOAT claim for prompt emission by comparing the Konus and GBM observations of GRB\,221009A with the broader sample of (nearly) all prompt GRBs identified since their discovery. The \textit{Fermi}-GBM paper \citep{grb221009a_lesage_fermi_gbm_2023} and the Konus-\textit{Wind} paper \citep{grb221009a_konus_2023} focus on the respective analyses of this burst and place it into context of the respective samples. We refer the reader to both papers for those details. This paper makes use of the analyses in both papers, and is intended to be complementary. Our sample compilation and input catalogs are explained in Section\,\ref{sec:catalogs}. The bright samples for fluence, peak flux, $E_{\rm iso}$ and L$_{\rm iso}$ are presented in Section\,\ref{sec:context}. The immediate implications of our work are explored in Section\,\ref{sec:discussion}, and we conclude with Section\,\ref{sec:conclusion}.

\section{Sample} \label{sec:catalogs}
Numerous GRB monitors have been launched to study these events \citep{universe8070373}. The monitors of focus for this work and key metrics are given in Table\,\ref{tab:instruments}. For bright bursts, observed space-time volume can be represented as a continuous full-sky equivalent value, i.e. number of $4\pi$-years. The maximal value is $\sim55$, from the discovery of the first GRB in 1967 until the end of our sample on March 7th, 2023; however, not all data is publicly available. Notable gaps in coverage occur from the discovery of GRBs in 1967 until the start of our \textit{Vela} sample and from the end of our \textit{Vela} sample until the launch of \textit{Pioneer Venus Orbiter} (\textit{PVO}), limiting the available maximal value to $\sim48$. Altogether, our data from \textit{Vela}, \textit{PVO}, the Burst And Transient Source Experiment (BATSE), Konus-\textit{Wind}, and \textit{Fermi}-GBM provide a $4\pi$-year equivalent coverage of 44.3\,years, which is $\sim92$\% of the possible public total. An additional literature search, described below, has increased this fraction. We thus expect our fluence and peak flux coverage to be largely complete.

\begin{table}[]
\begin{tabular}{r|lllll}
Instrument                      & \textit{Vela}     & \textit{PVO}      & BATSE   & Konus-\textit{Wind}  & \textit{Fermi}-GBM   \\ \hline
Start Date                      & 690703   & 780914   & 910421  & 941112      & 080714      \\
End Date                        & 730610   & 921004   & 000513  & 230307*     & 230307*     \\
Calendar Observing Years        & 3.9     & 13.3    & 9.7    & 28.3       & 14.6        \\
All-sky Observing Fraction      & 100\%    & 100\%    & 67\%    & 100\%       & 70\%        \\
Livetime Fraction               & 100\%    & 90\%     & 73\%    & 90\%        & 85\%        \\
$4\pi$-year Equivalent Coverage & 3.9      & 11.9     & 4.7     & 25.5        & 8.8         \\
Fluence Reporting Range {[}keV{]}          & 300-1,500 & 100-2,000 & 20-2,000 & 20-10,000    & 10-1,000     \\
Peak Flux Interval {[}s{]}      &          & 0.250    & 2.048   & 0.064       & 1.024       \\
GRB Sample Size                    & 20       & 318      & 2704    & $\sim$3,500 & $\sim$3,500
\end{tabular}
\caption{Basic properties and $4\pi$-year equivalent coverage for GRB monitors whose data was studied in detail for this catalog. The BATSE observing fraction is determined from orbital height and livetime inferred from \citet{2003AIPC..662..176H}. \textit{Vela} is assumed to have 100\% coverage. Other values are taken from the cited references for each instrument. Note that the End Dates for Konus and GBM refer to the end of our sample, as do the total number of GRBs; both are still observing at the time of publication.
}\label{tab:instruments} 
\end{table}

While not utilized to construct our GRB sample, we highlight the importance of the \textit{Neil Gehrels Swift Observatory} (\textit{Swift}, hereafter) for this work \citep{gehrels2004swift}. Its arcminute-scale localizations by the Burst Alert Telescope \citep[BAT,][]{Barthelmy05}, prompt follow-up with the narrow-field instruments XRT and UVOT \citep{burrows2005swift,roming2005swift}, and immediate reporting to the follow-up community has been critical for construction of the redshift sample of GRBs. For bursts with reported redshift and broadband spectral observations, \textit{Swift} localizations with Konus-\textit{Wind} and \textit{Fermi}-GBM spectral coverage provide the vast majority of the total sample.

\subsection{Input Sample}\label{sec:input}
While GRBs were discovered in 1967, quantitative study of their brightness began with GRB 690703, the first GRB observed with the \textit{Vela} 5A and 5B satellite pair launched in May of that year. Our \textit{Vela} sample begins with that burst and ends in 1973, the last year with publicly reported data. Fluence values are taken from \citet{1974ApJ...188L...1S}; peak flux measurements are not available. \textit{PVO} observed for 14\,years with all-sky coverage and high livetime \citep{klebesadel1980pioneer,fenimore_pvo_overview}, covering a quarter of the total possible $4\pi$-year sample. The \textit{PVO} fluence and peak flux values are taken from \citet{pvo_fenimore_2023}, which is an updated version using additional checks from the data presented in \citet{fenimore1993intrinsic}. The most sensitive GRB monitor ever flown is \textit{CGRO}-BATSE. The BATSE fluence and peak flux values are compiled in the 5B Spectral Catalog \citep{goldstein2013batse}, available on HEASARC\footnote{\url{https://heasarc.gsfc.nasa.gov/W3Browse/cgro/bat5bgrbsp.html}}.
 
With nearly 28 years of full sky observing with high livetime, Konus-\textit{Wind} \citep{aptekar1995konus} alone covers more than half of the total sample of bright bursts. A search for the highest fluence and peak flux bursts in the Konus sample was performed for this work and is expected to be complete in fluence down to a few$\times10^{-4}$\,erg\,cm$^{-2}$. The E$_{\rm iso}$ and L$_{\rm iso}$ samples for Konus are compiled in two publications \citep{tsvetkova2017konus,tsvetkova2021konus} with updates through March 2023 for this work. 

Our last considered instrument is the \textit{Fermi}-GBM. The $4\pi$-year equivalent coverage of GBM is limited by the particle activity and Earth blockage of the sky inherent to a Low Earth Orbit (which also affects BATSE). The GBM observations are additionally critical for proper use of the full \textit{Vela}, \textit{PVO}, and BATSE GRB samples. GBM has the widest energy coverage of any GRB monitor and the GBM 10 Year Spectral Catalog \citep{poolakkil2021fermi} is complete with respect to the on-board trigger catalog \citep{von2020fourth}, both of which are available on HEASARC\footnote{\url{https://heasarc.gsfc.nasa.gov/W3Browse/fermi/fermigtrig.html},\url{https://heasarc.gsfc.nasa.gov/W3Browse/fermi/fermigbrst.html}}. These together allow for the standardization of fluence and peak flux measurements in our various input instruments, described in the next section.

Lastly, to capture all identified bright bursts, we utilized GRBCAT\footnote{\url{https://heasarc.gsfc.nasa.gov/grbcat/}}, the BeppoSAX GRB Spectral Catalog \citep{guidorzi2011spectral}, a search of NASA ADS for known and forgotten bright bursts, and searched for recent bright bursts reported in publications or initial results in GCN circulars. All bright Konus-\textit{Venera} GRBs reported \citep{1981Ap&SS..80....3M} are contained in the \textit{PVO} catalog. It is feasible that some bright bursts escaped our searches. Notably, there is no reported GRB catalog from \textit{GINGA}. However, with the \textit{Vela} satellites and successors we are confident that any burst of sufficient brightness that may affect our conclusions would certainly be known, and we are confident in the claims that follow.

\subsection{Standardizing the Sample}\label{sec:standardize}
Owing to different instrument designs and analysis decisions our input catalogs report brightness measurements integrated over different energy ranges. Comparison between instruments requires conversion to a uniform energy range. In order to account for nearly all emission, we set this uniform energy range to the standard bolometric range of 1\,keV - 10\,MeV \citep[e.g.][]{racusin2009jet,tsvetkova2017konus}. For intrinsic measures these values are $k$-corrected \citep{bloom2001prompt}. Although this will miss significant high energy emission in a small subset of bursts \citep[see][for an application of an even broader energy range]{AguiFernandez2023}, it is better matched to the observing range of all considered instruments. 

For Konus-\textit{Wind} we utilize the standard reported values of 20\,keV-10\,MeV for fluence and peak flux, which are sufficiently close to bolometric values, and the catalog-reported $k$-corrected 1\,keV-10\,MeV values for E$_{\rm iso}$ and L$_{\rm iso}$ \citep{tsvetkova2017konus}. For all other instruments, we convert measurements to the bolometric energy band of 1\,keV-10\,MeV; for intrinsic measures this is defined in the rest frame and accounts for cosmological $k$-correction. For GBM and BATSE bursts best-fit by spectra with curvature \citep[as determined through the standard catalog methods,][]{goldstein2013batse,poolakkil2021fermi} we directly calculate bolometric energetics by sampling parameter value distributions for proper, asymmetric error propagation. 

For \textit{PVO} and \textit{Vela} bursts we cannot directly calculate bolometric energetics. For GBM and BATSE bursts best-fit by a power-law we cannot accurately extrapolate to bolometric brightness due to lack of determination of spectral curvature. For each instrument, we utilize the GBM sample \citep{poolakkil2021fermi} to determine the scaling distribution from the initial energy range to the bolometric energy range. This is determined separately for peak flux and fluence, owing to the different hardness in peak intervals verse time-integrated spectra. \textit{PVO} values are scaled from the 50-300\,keV peak flux and fluence values as they were determined using the modern BAND function \citep{band1993batse}. For determining the scaling distributions to apply to \textit{Vela} we exclude GBM bursts with $E_{peak}$ less than 300\,keV as similar bursts are unlikely to trigger the \textit{Vela} instruments due to the higher low-energy threshold. Further details of this procedure are described in Appendix\,\ref{app:peak_flux_scalings}.

Given the sharply peaked pulse structure of GRBs, shorter peak flux intervals will correlate with high inferred peak flux values. In Appendix\,\ref{app:peak_flux_scalings} we show that peak flux values will increase by $\sim15$\% for each step of two towards shorter intervals. For all L$_{\rm iso}$ measures we have converted the native 0.064\,s interval for Konus, 0.250\,s interval for \textit{PVO}, and 2.048\,s interval for BATSE to the 1.024\,s interval for GBM, matching the measured timescale for GRB\,221009A. An inversion of this procedure is used for logN-logP comparisons in Section\,\ref{sec:rarity}. %This is used for a small portion of the peak flux analysis and is applied to all reported L$_{\rm iso}$ values. %\SL{(please note that this is not 100\% statistically correct for \textit{PVO} (because 1.024 is not an exact multiple of 0.25) or BATSE (because you are going from a higher timescale to a lower one))}. 

Uncertainties are fully calculated throughout this paper. For E$_{\rm iso}$ and L$_{\rm iso}$ the typical $1-\sigma$ fractional uncertainty is $\lesssim10$\%. For fluence and peak flux the corresponding numbers are 35\% and 45\%, respectively, which are larger due to the application of our scaling methods to some bursts (detailed in Appendix\,\ref{app:bolo_scalings}, \ref{app:peak_flux_scalings}), which are only rarely necessary for the isotropic-equivalent energetics calculations. The uncertainties do not affect any of our conclusions and are therefore omitted for brevity.

To check inter-calibration uncertainty between instruments, fluence was compared for bursts seen by two instruments, showing average overall agreement within $\sim20$\%. Additional confirmation that this approach is reasonable is given by the logN-logS agreement in the next section. To match the sample to the long GRB class of GRB\,221009A, we exclude bursts that are obviously of the cosmological short class (durations under 2\,s). We additionally exclude bursts which are known to be magnetar giant flares, as they originate from a distinct physical origin. Thus, the samples below should be mostly comprised of long GRBs arising from collapsars. For the intrinsic energetic figures, some short bursts are shown for comparison.

% \subsection{Additional Bursts}
% \begin{itemize}
%     \item GRB 830801 - 2E-3 \url{https://adsabs.harvard.edu/full/1987SvAL...13..444K}, 
%     \url{https://articles.adsabs.harvard.edu/pdf/1986SvAL...12..315K}
%     \item \textcolor{red}{Ed: there is GRB 830801. It is a very bright burst, but there is a few month gap in the \textit{PVO} sample. }
%     \item 910402 - \url{https://arxiv.org/pdf/astro-ph/0005293.pdf}
%     \item PHEBUS-\textit{GRANAT}, 'GRB 940703A' ~5E-4 (not 1E-3) \url{https://aip.scitation.org/doi/pdf/10.1063/1.55335}
%     \item \SL{GRB 990510 (?)}
% \end{itemize}

% \textcolor{red}{Distributions built up from 14 years of \textit{PVO} (per Ed)
% and 28 years of Konus-WIND, for 42 of 55 years covered. Invesigate the remaining 13 years for any burst that could have a brighter fluence and/or flux than GRB 221009A. For 730610 - 780420 and 921008 to 941101 these will be GRBCAT.}

\section{GRB\,221009A in context} \label{sec:context}
Peak flux and fluence values for GRB\,221009A are taken from Konus \citep{grb221009a_konus_2023} and GBM \citep{grb221009a_lesage_fermi_gbm_2023}. Both observations are non-standard, given the unprecedented brightness of this event. The Konus and GBM teams worked in isolation before comparing values, allowing for independent checks on reconstruction accuracy. We additionally compare with the brightness measures from \textit{INSIGHT}-HXMT and \textit{GECAM-C}, reported in \cite{grb221009_gecam_2023}. The GBM numbers presented here are preliminary; however, due to general agreement with other values of the brightness of prompt emission of GRB\,221009A the values are suitably robust for our purposes here. For intrinsic energetics we use the redshift of 0.151 \citep{2022GCN.32648....1D,Malesani2023redshift} and a typical cosmology\footnote{That is, the default flat universe with H$_0=69.6$ and $\Omega_m=0.286$ from \url{https://astro.ucla.edu/~wright/CosmoCalc.html}} giving a luminosity distance of 724\,Mpc. When a given burst is identified by more than one facility we use the highest brightness value reported.

While this paper was under review, GRB 230307A was detected by GBM, Konus, and other instruments \citep{gcn_gbm_230307a,230307a_gcn_konus}. With the initial results this burst has the second highest energy fluence ever reported. The initial results for this burst are included for the fluence and peak flux discussions below, but not in the intrinsic energetics section (as a redshift is currently not known).

%This gives an $E_{\rm iso}$ and $L_{\rm iso}$ of \textcolor{red}{values}

\subsection{Fluence}\label{sec:fluence}
\begin{figure*}[!h]
	\centering
	\includegraphics[width=1.0\textwidth]{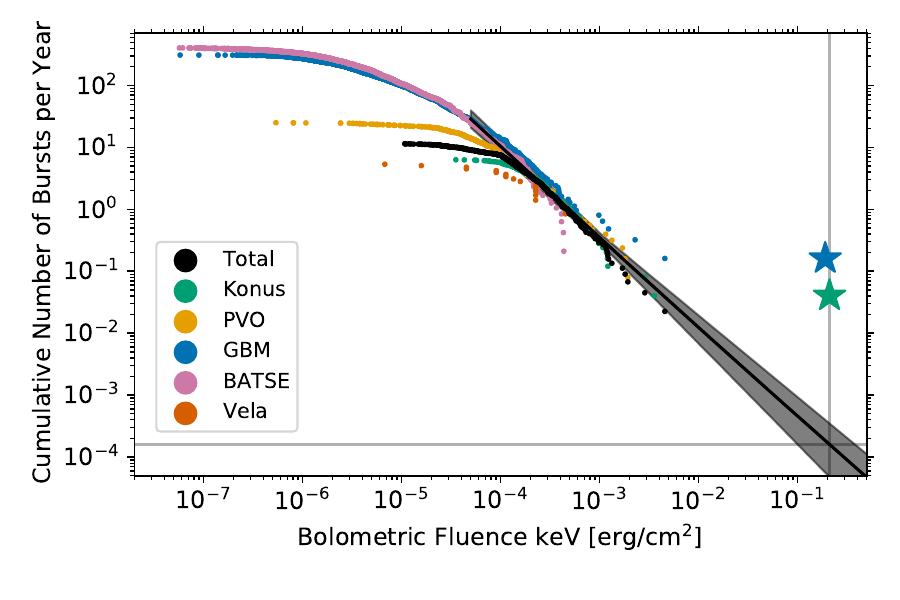}
    \caption{Points indicate the annualized logN-logS distributions for GBM, \textit{PVO}, BATSE, Konus, and \textit{Vela}. A merged total sample is presented with a fit (from Section\,\ref{sec:rarity}) to the combined logN-logS distribution with index measured as $-1.47\pm0.15$ (90\% confidence interval). GRB\,221009A stands alone, with the Konus and GBM measurements of this burst denoted by stars. A vertical line maps the observed fluence of GRB\,221009A to the power-law extrapolation, and the horizontal line marks the inverse recurrence rate of events this bright. %A burst of this fluence arrives at Earth approximately every $6,200-8,900$ years.
    }
    \label{fig:fluence}
\end{figure*}

The fluence S of GRB\,221009A is $0.21\pm0.02$\,erg\,cm$^{-2}$ as measured by Konus-\textit{Wind} \citep{grb221009a_konus_2023} and $\sim0.19$\,erg\,cm$^{-2}$ as measured by \textit{Fermi}-GBM \citep{grb221009a_lesage_fermi_gbm_2023}. These values are in agreement with the \textit{INSIGHT}-HXMT and \textit{GECAM-C} bolometric fluence of $0.224\pm0.002$ \citep{grb221009_gecam_2023}. logN-logS is the cumulative number N events above a given fluence S; GRB\,221009A is compared against our annualized logN-logS distributions from our considered instruments in Figure\,\ref{fig:fluence}. Our samples of interest show broad agreement, noting the significant uncertainty due to low counts at particularly high fluence. In this regime, truncation due to instrumental limitations may also be significant.

\begin{table}[h]
\begin{tabular}{llllll}
GRB Name    & Duration & Fluence                   & Energy Range & Instrument           & Reference                                                                             \\ \hline
            & {[}s{]}  & [erg\,cm$^{-2}$] & {[}keV{]}            &                      &                                                                                       \\ \hline
GRB 221009A & 600   & 0.21                      & 1-10,000             & Konus, GBM           & \citet{grb221009a_konus_2023,grb221009a_lesage_fermi_gbm_2023} \\
GRB 230307A & 200   & $4.56\times10^{-3}$                      & 1-10,000             & GBM, Konus           & \citet{gcn_gbm_230307a,230307a_gcn_konus} \\
GRB 130427A &  62        & $2.86\times10^{-3}$                  & 20-10,000            & Konus, GBM                &    \citet{tsvetkova2017konus,poolakkil2021fermi}                                                                                   \\
GRB 840304  & 1000     & $\sim2.8\times10^{-3}$                  & 1-10,000             & \textit{PVO}                  & \citet{klebesadel1984unusual,chuang1990identification}                           \\
GRB 830801  & 30        & $>2.00\times10^{-3}$                  & 30-7,500             & SIGNE 2 \textit{MP9}          & \citet{kuznetsov1987analysis}                                        \\
GRB 920212  & 14       & $1.93\times10^{-3}$                  & 1-10,000             & \textit{PVO}                  & \citet{pvo_fenimore_2023}, This work                             \\
GRB 900808  & 7.2      & $1.81\times10^{-3}$                  & 1-10,000             & \textit{PVO}                  & \citet{pvo_fenimore_2023}, This work                             \\
GRB 940703A & 31.4     & $1.60\times10^{-3}$                  & 100-10,000           & PHEBUS-\textit{GRANAT} & \citet{barat1998phebus}                                              \\
GRB 811016  & 13.2     & $1.33\times10^{-3}$                  & 1-10,000             & \textit{PVO}                  & \citet{pvo_fenimore_2023}, This work                             \\
GRB 160625B & 680      & $1.23\times10^{-3}$                  & 1-10,000             & GBM, Konus           & \citet{poolakkil2021fermi}, This work                                \\
GRB 180914B &  150        & $1.21\times10^{-3}$                  & 20-10,000            & Konus                &                                                              \citet{2018GCN.23240....1F}                         \\
GRB 140219A & 18         & $1.20\times10^{-3}$                  & 20-10,000            & Konus                &                                                             \citet{2014GCN.15870....1G}                          \\
GRB 160821A & 47         & $1.17\times10^{-3}$                  & 20-10,000            & Konus, GBM                &   \citet{2016GCN.19842....1K,poolakkil2021fermi}                 \\
GRB 911027  & 111      & $1.15\times10^{-3}$                  & 1-10,000             & \textit{PVO}                  & \citet{pvo_fenimore_2023}, This work                             \\
GRB 710630  & $\sim$7  & $1.13\times10^{-3}$                  & 1-10,000    & \textit{Vela}                 & \citet{1974ApJ...188L...1S}, This work                               \\
GRB 910402  & 35.9     & $1.11\times10^{-3}$                  & 100-10,000           & PHEBUS-\textit{GRANAT} & \citet{barat1998phebus}                                              \\
GRB 021206  & 5.2        & $1.08\times10^{-3}$                  & 20-10,000            & Konus                &    This Work                                                                                 
\end{tabular}
\caption{GRBs with fluence $>10^{-3}$\,erg\,s$^{-1}$\,cm$^{-2}$. The brightest measurements for a given burst are reported. GRB\,940703A and GRB\,910402 were identified by BATSE, but we use the higher values reported by PHEBUS-\textit{GRANAT} measures for the same bursts. GRB\,830801, perhaps the third highest fluence ever, was measured by SIGNE 2 \textit{MP9}. The value for GRB\,840304 is taken from the dedicated analysis on this burst. References contain more details on individual bursts. Durations here are not a uniformly measured quantity (estimated from lightcurve, T$_{90}$, T$_{100}$) and are only intended to be approximate. %\textcolor{purple}{Add correct Konus references, duration (EB: I can point to circulars, if appropriate)} %References for each burst contain more details.
}\label{tab:fluence}
\end{table}

Table\,\ref{tab:fluence} contains the brightest bursts in our sample, including bursts from additional instruments beyond those considered in our main sample. Of the 15 other bursts with a bolometric fluence in excess of $10^{-3}$\,erg\,s$^{-1}$\,cm$^{-2}$, only 2 have durations comparable or longer than GRB\,221009A.

\subsection{Peak Flux}\label{sec:flux}
\begin{figure*}[!h]
	\centering
	\includegraphics[width=1.0\textwidth]{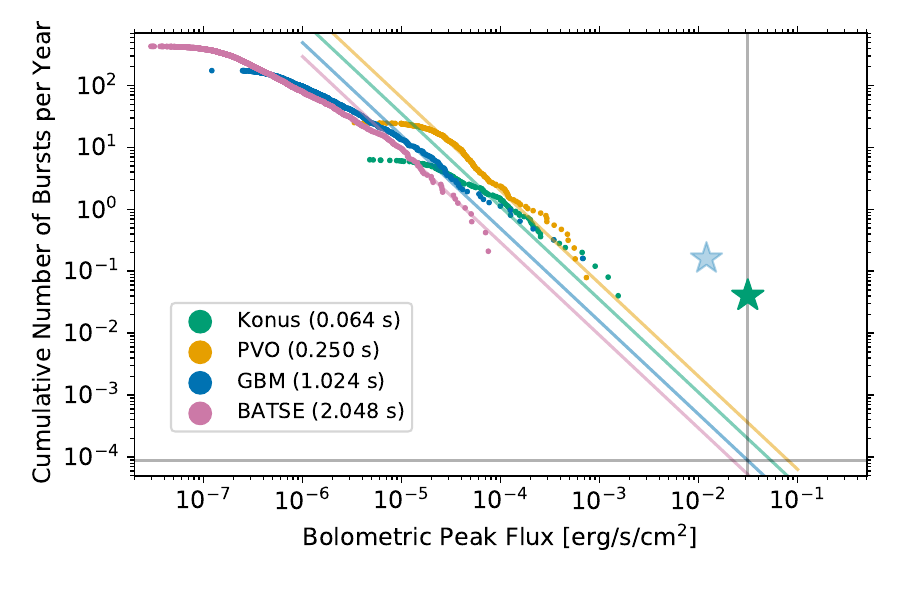}
    \caption{Points indicate the annualized logN-logP distributions for GBM, \textit{PVO}, BATSE, and Konus, with peak flux interval indicated in the legend. A power-law with fixed index of -3/2 is shown for each separate burst sample. The GBM and BATSE distributions are fit above 1$\times10^{-5}$\,erg\,s$^{-1}$\,cm$^{-2}$; the Konus and \textit{PVO} distributions are fit above 5$\times10^{-5}$\,erg\,s$^{-1}$\,cm$^{-2}$. GRB\,221009A measures are shown with stars and again it is a significant outlier. The GBM measure of GRB\,221009A is faded to indicate it may be an underestimate.
    }
    \label{fig:flux}
\end{figure*}

For GRB\,221009A the peak flux P measured over a 1\,s timescale by Konus-\textit{Wind} is 0.031$\pm$0.005\,erg\,s$^{-1}$\,cm$^{-2}$, with temporal precision limited by the return to lower resolution data. The preliminary 1.024\,s peak flux as measured by \textit{Fermi}-GBM of this burst is 0.01\,erg\,s$^{-1}$\,cm$^{-2}$. The \textit{INSIGHT}-HXMT and \textit{GECAM-C} 1\,s peak flux value of $0.0172\pm0.0003$ is consistent \citep{grb221009_gecam_2023}. We follow our procedure of using the highest reported value for a given measure. These values is compared against the annual logN-logP distributions from our instruments in Figure\,\ref{fig:flux}.

\begin{table}[h]
\begin{tabular}{lllll}
GRB Name    & Interval [s] & Peak   Flux       & Instrument           & Reference                                                      \\ \
  & [s] & [erg\,s$^{-1}$\,cm$^{-2}$]     &          &                                                      \\ \hline
GRB 221009A & 1.024        & 0.031              & Konus, GBM           & \citet{grb221009a_konus_2023,grb221009a_lesage_fermi_gbm_2023} \\
%GRB 020214  & 0.064        & 1.54$\times10^{-3}$     & Konus      & This work     removed by Dmitry S, there was a typo in my table                                               \\
GRB 140219A & 0.064        & 1.22$\times10^{-3}$          & Konus                & This work                                                      \\
GRB 110918A & 0.064        & 9.02$\times10^{-4}$          & Konus                & \citet{Frederiks_2013ApJ_GRB110918A,tsvetkova2017konus}                                                     \\
GRB 920212  & 0.25         & 7.36$\times10^{-4}$          & \textit{PVO}                  & \citet{pvo_fenimore_2023}                                      \\
GRB 130427A & 0.064        & 6.81$\times10^{-4}$          & Konus, GBM           & \citet{tsvetkova2017konus,poolakkil2021fermi}                  \\
GRB 230307A & 1.024        & 6.71$\times10^{-4}$          & GBM, Konus           & \citet{gcn_gbm_230307a,230307a_gcn_konus}                  \\
GRB 830801  & 1            & $\gtrsim$6.67$\times10^{-4}$ & SIGNE 2 \textit{MP9}          & Inferred from \citet{kuznetsov1987analysis}                     \\
GRB 900808  & 0.25         & 5.67$\times10^{-4}$          & \textit{PVO}                  & \citet{pvo_fenimore_2023}                                      \\
GRB 890923  & 0.25         & 5.54$\times10^{-4}$          & \textit{PVO}                  & \citet{pvo_fenimore_2023}                                      \\
GRB 811016  & 0.25         & 4.81$\times10^{-4}$          & \textit{PVO}                  & \citet{pvo_fenimore_2023}                                      \\
GRB 911226  & 0.25         & 4.73$\times10^{-4}$          & \textit{PVO}                  & \citet{pvo_fenimore_2023}                                      \\
GRB 021206  & 0.064        & 4.50$\times10^{-4}$          & Konus                & This work                                                      \\
GRB 160625B & 1.024        & 2.13$\times10^{-4}$          & GBM, Konus           & \citet{poolakkil2021fermi,tsvetkova2017konus}                  \\
GRB 131014A & 1.024        & 1.57$\times10^{-4}$          & GBM, Konus           & \citet{poolakkil2021fermi,tsvetkova2017konus}                  \\
GRB 910402  & 1.62         & 1.47$\times10^{-4}$          & PHEBUS-\textit{GRANAT}, BATSE & \citet{barat1998phebus}                                        \\
GRB 940703  & 7.64         & 1.37$\times10^{-4}$          & PHEBUS-\textit{GRANAT}, BATSE & \citet{barat1998phebus}                                        \\
GRB 171227A & 1.024        & 1.26$\times10^{-4}$          & GBM, Konus           & This work                                                      \\
GRB 160821A & 1.024        & 1.25$\times10^{-4}$          & GBM, Konus           & \citet{poolakkil2021fermi,tsvetkova2017konus}                                   
\end{tabular}
\caption{The highest peak flux GRBs from our input samples, selecting some bursts with the highest values in each instrument sample. GRB\,940703A and GRB\,910402 were identified by BATSE, but we use the higher PHEBUS-\textit{GRANAT} values for the same bursts. As no peak flux is reported for GRB\,830801 we estimate a lower limit from \cite{kuznetsov1987analysis} noting that most of the energy flux occurs within a $\sim3$\,s interval, peaked at T0+0.5 to T0+1.5\,s. No claimed peak flux is within an order of magnitude of GRB\,221009A.
}\label{tab:flux}
\end{table}

The peak flux of GRB\,221009A is most directly compared with the GBM values, given the 1.024\,s peak interval. As expected, the 2.048\,s peak flux distribution from BATSE has systematically lower values than the 1.024\,s distribution from GBM, which is lower than the 0.064\,s distribution from Konus. With proper selection of considered events and analysis of BATSE data not available in the BATSE catalogs, the \textit{PVO} and BATSE logN-logP distributions show strong agreement \citep{fenimore1993intrinsic}; however, our 0.250\,s \textit{PVO} distribution is anomalously high. The origin of this is not well understood but may arise as the cataloged peak flux values are in photons, not ergs, which differs from the rest of our input samples. This issue does not affect our conclusions: GRB\,221009A is again an obvious outlier with no burst within an order of magnitude. The brightest individual peak flux bursts are reported in Table\,\ref{tab:flux}.

\subsection{Total Intrinsic Energy}\label{sec:eiso}
\begin{figure*}[!h]
	\centering
	\includegraphics[width=1.0\textwidth]{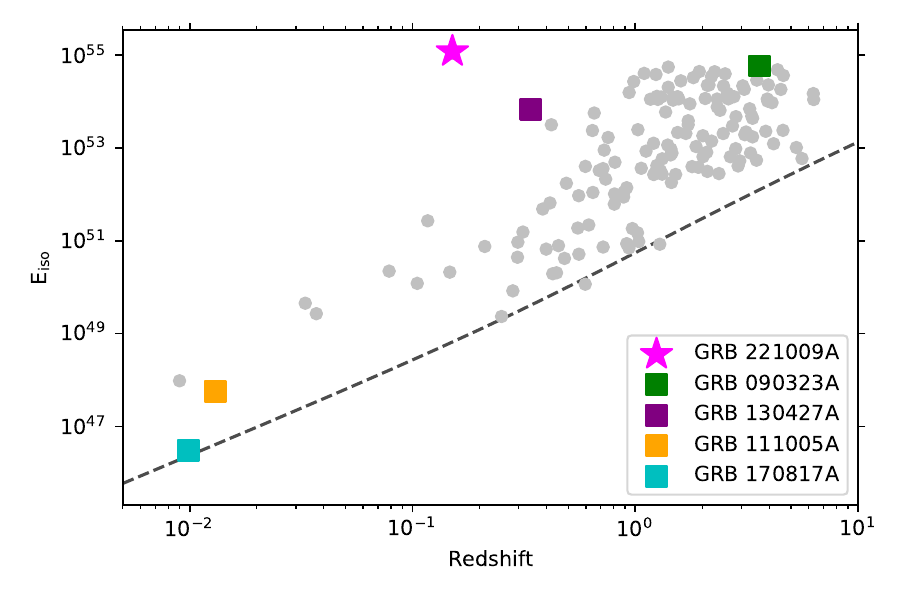}
    \caption{Bolometric, $k$-corrected E$_{\rm iso}$ for nearly 150 GRBs compiled from the literature \citep{GW170817-GRB170817A,tsvetkova2017konus} with additional bright and faint GRBs compiled following those works and for future analyses. The dashed line is an approximate, empirical detection threshold for GRBs as a function of redshift. Extreme GRBs are highlighted: GRB\,130427A the previous fluence record holder, GRB\,090323A the previous E$_{\rm iso}$ record holder, GRB\,111005A the lowest known $E_{\rm iso}$ for a collapsar, and GRB\,170817A the lowest known $E_{\rm iso}$ overall. GRB\,221009A is the record holder.
% \textcolor{purple}{Konus ultra-long GRBs?}
    }
    \label{fig:eiso}
\end{figure*}

GRB\,221009A is obviously the BOAT as measured by prompt gamma-ray fluence, but the nearby distance and instrumental issues in the large GRB monitors due to burst brightness leaves the question of how the total intrinsic energy, E$_{\rm iso}$, compares to the broader sample. Of the $\sim400$\,GRBs with measured intrinsic energetics, those in \citet{tsvetkova2017konus}, \citet{tsvetkova2021konus}, \citet{GW170817-GRB170817A}, and additional bursts compiled here, GRB\,221009A is also the E$_{\rm iso}$ record holder. The Konus-\textit{Wind} measurement is $\sim$1.2$\times10^{55}$\,erg while the \textit{Fermi}-GBM measurement is $\sim$1.0$\times10^{55}$\,erg, and \cite{grb221009_gecam_2023} even derive $\sim$1.5$\times10^{55}$\,erg from \textit{INSIGHT}-HXMT and \textit{GECAM-C} data. We have compiled a large E$_{\rm iso}$ sample, focusing on those with the highest measured values, shown in Figure\,\ref{fig:eiso}. The closest bursts are only $\sim50$\% the value of GRB\,221009A, as shown in Table\,\ref{tab:eiso}.

\begin{table}[h]
\begin{tabular}{llllll}
GRB Name    & Redshift & Duration     & Eiso    & Instrument & Reference                                                                             \\ \hline
  & &  [s]     & [erg]     & &                                                                         \\ \hline
GRB 221009A & 0.151    & 600     & $\sim1.2\times10^{55}$ & Konus, GBM & \citet{grb221009a_konus_2023,grb221009a_lesage_fermi_gbm_2023} \\
GRB 090323  & 3.6      & 130  & $5.81\times10^{54}$ & Konus, GBM & \citet{tsvetkova2017konus}                                           \\
GRB 160625B & 1.406    & 680 & $5.50\times10^{54}$ & GBM, Konus & \citet{GW170817-GRB170817A,tsvetkova2017konus}                       \\
GRB 080916C & 4.35     & 63  & $4.82\times10^{54}$ & Konus, GBM & \citet{tsvetkova2017konus,GW170817-GRB170817A}                       \\
GRB 210619B & 1.937    &  52       & $4.41\times10^{54}$ & Konus      & This work                                                                             \\
GRB 130505A & 2.27     &   32      & $4.37\times10^{54}$ & Konus      & \citet{tsvetkova2017konus}                                           \\
GRB 180914B & 1.096    &   150      & $4.03\times10^{54}$ & Konus      & This work                                                                             \\
GRB 170214A & 2.53     &  150       & $3.94\times10^{54}$ & Konus      & This work                                                                             \\
GRB 130907A & 1.238    &  210       & $3.82\times10^{54}$ & Konus      & \citet{tsvetkova2017konus}                                           \\
GRB 220101A & 4.618    & 240        & $3.64\times10^{54}$ & Konus, GBM      & This work                                                                             \\
GRB 120624B & 2.1974   & 270 & $3.45\times10^{54}$ & GBM, Konus & \citet{GW170817-GRB170817A,tsvetkova2017konus}                       \\
GRB 090902B & 1.822    & 19  & $3.26\times10^{54}$ & GBM        & \citet{GW170817-GRB170817A}                                          \\
GRB 170405A & 3.51     & 80  & $2.89\times10^{54}$ & Konus, GBM & This work, \citet{GW170817-GRB170817A}                               \\
GRB 990123  & 1.6004   &   110      & $2.78\times10^{54}$ & Konus, BATSE      & \citet{tsvetkova2017konus}                                           \\
GRB 110918A & 0.984    &   95      & $2.69\times10^{54}$ & Konus      & \citet{tsvetkova2017konus}                                           \\
% GRB 140419A & 3.956    &   58      & 2.28E+54 & Konus      & \citet{tsvetkova2017konus}                                           \\
% GRB 230204B & 2.142    & 216 & 2.27E+54 & GBM        & This Work                                                                             \\
% GRB 090926A & 2.1062   & 13.8   & 2.23E+54 & GBM, Konus & \citet{tsvetkova2017konus}                                           \\
% GRB 080607  & 3.0363   &   47      & 2.17E+54 & Konus      & \citet{tsvetkova2017konus}                                           \\
% GRB 130518A & 2.488    & 49  & 2.16E+54 & Konus, GBM & \citet{tsvetkova2017konus}                                           \\
% GRB 120711A & 1.405    & 44  & 2.03E+54 & Konus, GBM & \citet{tsvetkova2017konus}                                          
\end{tabular}
\caption{GRBs with E$_{\rm iso} >2.5\times10^{54}$\,erg. Bursts prior to the launch of \textit{Swift} were included but only GRB\,990123 meets our threshold. GRB\,221009A is the highest by nearly a factor of 2. Durations here are not a uniformly measured quantity (estimated from lightcurve, T$_{90}$, T$_{100}$) and are only intended to be approximate.}
\label{tab:eiso}
\end{table}

\subsection{Peak Isotropic-Equivalent Luminosity}
\begin{figure*}[!h]
	\centering
	\includegraphics[width=1.0\textwidth]{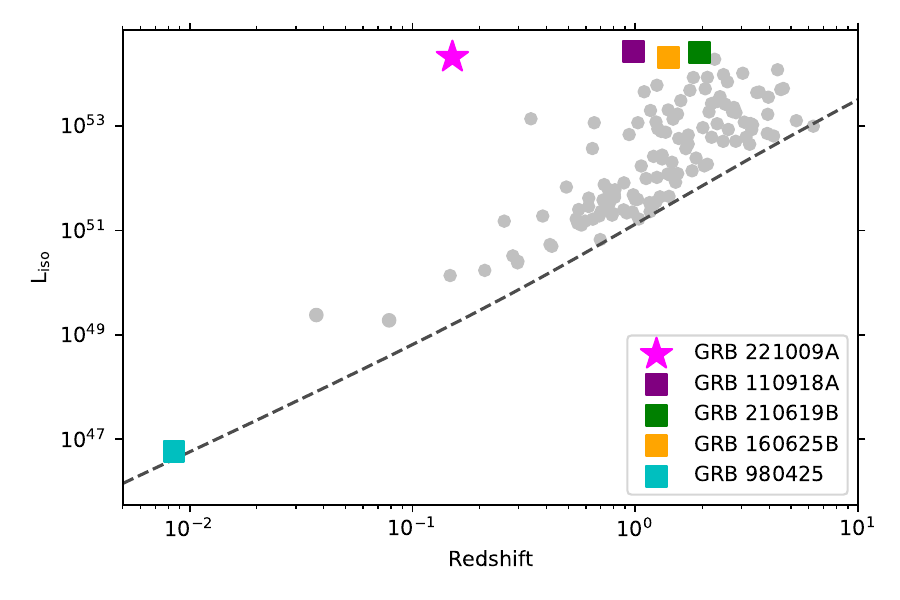}
    \caption{Bolometric, $k$-corrected L$_{\rm iso}$ for a nearly 100 GRBs compiled from the literature \citep{GW170817-GRB170817A,tsvetkova2017konus} with additional bright GRBs compiled following those works. The dashed line is an approximate, empirical detection threshold for GRBs as a function of redshift. Extreme GRBs are highlighted: GRBs 110918A and 210619B have higher L$_{\rm iso}$ than GRB\,221009A. GRB\,160625B is marked as it is an analog of GRB\,221009A, discussed in Section\,\ref{sec:analogs}. We also show GRB\,980425, as measured by BATSE, as one of the lowest known $L_{\rm iso}$ values in a GRB. GRB\,221009A is \textit{not} the record holder. 
    }
    \label{fig:liso}
\end{figure*}

The peak luminosity of GRB\,221009A is measured by Konus-\textit{Wind} to be $\sim2.1\times10^{54}$\,erg\,s$^{-1}$ \citep{grb221009a_konus_2023} and by \textit{Fermi}-GBM to be $\sim1.0\times10^{54}$\,erg\,s$^{-1}$ over the 1.024\,s interval. The other L$_{\rm iso}$ values we compare with are taken over the 1.024\,s peak interval. For GBM-detected GRBs this is the reported value; for others we scale the reported value as described in Appendix\,\ref{app:peak_flux_scalings}. GRB\,221009A has an extreme but not record L$_{\rm iso}$, being at the $\sim99$th percentile of GRBs. The highest L$_{\rm iso}$ bursts are reported in Table\,\ref{tab:liso} and the broader sample shown in Figure\,\ref{fig:liso}.

\begin{table}[h]
\begin{tabular}{lllll}
GRB Name    & Redshift & Liso     & Instrument & Reference                                                                             \\ \hline
  & & [erg/s]  & &                                                                         \\ \hline
GRB 110918A & 0.984    & $2.70\times10^{54}$ & Konus      & \citet{tsvetkova2017konus}                                           \\
GRB 210619B & 1.937    & $2.53\times10^{54}$ & Konus, GBM & This work                                                                             \\
GRB 221009A & 0.151    & $\sim2.1\times10^{54}$ & Konus, GBM & \citet{grb221009a_konus_2023,grb221009a_lesage_fermi_gbm_2023} \\
GRB 160625B & 1.406    & $2.04\times10^{54}$ & GBM, Konus & \citet{tsvetkova2017konus,GW170817-GRB170817A}                       \\
GRB 130505A & 2.27     & $1.91\times10^{54}$ & Konus      & \citet{tsvetkova2017konus}                                           \\
GRB 080916C & 4.35     & $1.20\times10^{54}$ & Konus, GBM & \citet{tsvetkova2017konus,GW170817-GRB170817A}                       \\
GRB 080607  & 3.0363   & $1.03\times10^{54}$ & Konus      & \citet{tsvetkova2017konus}                                           \\
GRB 130518A & 2.488    & $9.67\times10^{53}$ & GBM, Konus & \citet{tsvetkova2017konus,GW170817-GRB170817A}                       \\
GRB 090926A & 2.1062   & $8.58\times10^{53}$ & GBM, Konus & \citet{tsvetkova2017konus,GW170817-GRB170817A}                       \\
GRB 090902B & 1.822    & $8.54\times10^{53}$ & GBM        & \citet{GW170817-GRB170817A}                                          \\
GRB 060121  & 4.6      & $7.58\times10^{53}$ & Konus      & \citet{tsvetkova2017konus}                                           \\
GRB 080721  & 2.591    & $7.05\times10^{53}$ & Konus      & \citet{tsvetkova2017konus}                                           \\
GRB 200829A & 1.25     & $6.06\times10^{53}$ & Konus, GBM & This work                                                                             \\
GRB 220101A & 4.618    & $5.27\times10^{53}$ & Konus, GBM & This work                                                                             \\
GRB 150403A & 2.06     & $5.18\times10^{53}$ & GBM, Konus & \citet{tsvetkova2017konus,GW170817-GRB170817A}                       \\
GRB 000131  & 4.5      & $5.01\times10^{53}$ & Konus      & \citet{tsvetkova2017konus}                                          
\end{tabular}
\caption{GRBs with L$_{\rm iso} >5\times10^{53}$\,erg\,s$^{-1}$. Bursts prior to the launch of \textit{Swift} were included but only GRB\,000131 meets our threshold. GRB\,221009A is the third highest identified.}
\label{tab:liso}
\end{table}

% \subsection{Beaming-corrected Energetics and Comparison}
% \textcolor{red}{I am not sure we need to do this. The Konus sample is likely fine and we are unlikely to improve upon it}

\newpage
\section{Discussion} \label{sec:discussion}
GRB\,221009A is exceptional. It is, by far, the highest fluence and peak flux burst ever identified at Earth. It is additionally the record holder of E$_{\rm iso}$. It is also one of the highest L$_{\rm iso}$ bursts ever identified, though a few are known to be more luminous. These conclusions were independently made by both the Konus Team \citep{grb221009a_konus_2023} and the GBM Team \citep{grb221009a_lesage_fermi_gbm_2023}, and we refer to reader to their respective papers for these conclusions and detailed properties of the burst itself. In what follows, we explore the conclusions that can be drawn by comparing GRB\,221009A with the total sample of the brightest GRBs identified thus far.

% \textcolor{purple}{Note from Eric: Dmitry, Dmitry, Stephen, Peter - please let me know if you want to reserve any of the conclusions for the Konus / GBM Team papers. I am happy to remove things} %\SL{($\leftarrow$ you mentioned Dmitry twice, did you mean to add a different name?) Alex: Dmitry Frederiks and Dmitry Svinkin.}

\subsection{Rarity}\label{sec:rarity}
GRB\,221009A is out of class by both fluence and peak flux by around two orders of magnitude (when accounting for different peak flux intervals).
Our fluence sample is largely complete for the observed sample. We are certain that no GRB with higher fluence than GRB\,221009A exists in the observed sample, as it rivals even Galactic magnetar giant flares \citep{1999AstL...25..635M,palmer2005giant,frederiks2007giant}. The peak flux distribution is reasonably complete, though to a lesser extent than the fluence sample. However, any burst brighter than GRB\,221009A in the 55\,years of observations would certainly have been noted, even if it occurred during intervals without publicly accessible data.

The brightest portion of both the fluence and peak flux cumulative distributions are expected to follow a -3/2 power-law \citep[e.g.][]{Meszaros+95logNlogS}. GRBs are detected beyond the regime of local structure; thus, for a GRB of a fixed intrinsic brightness distributed in a sensitive volume the recovered signals will have a cubic power corresponding to the spatial volume and a square root power for the inverse square law of intrinsic-to-observed brightness. This holds for distances sufficiently close to be approximated as Euclidean; for an expanding universe, the lower fluence or peak flux cumulative functions will be shallower (due to the additional (1+\textit{z}) term for luminosity vs. comoving distance). This also holds only on the scale where source evolution is negligible. Both are true at the distance of GRB\,221009A. The -3/2 scaling has been observationally confirmed for the high fluence part of the logN-logS and logN-logP distributions \citep{1981Ap&SS..80....3M,fenimore1993intrinsic,meegan1992spatial,von2020fourth}. The shallower index for faint bursts \citep{meegan1992spatial} combined with their isotropic distribution is how the cosmological origin of GRBs was inferred \citep{briggs1995batse}.

The most robust bright sample available is our combined, bolometric logN-logS distribution constructed from bursts reported by \textit{Vela}, \textit{PVO}, BATSE, Konus-\textit{Wind}, and \textit{Fermi}-GBM (neglecting bursts seen only in other instruments), which has a value of 44.3 for effective $4\pi$-year coverage. Assuming a -3/2 power-law index and fitting the scale directly to the median fluence values for bursts above $3\times10^{-4}$\,erg\,s$^{-1}$\,cm$^{-2}$, where our sample should be largely complete, gives a CDF annual rate of GRBs above a given fluence S as R$_{\rm GRB}($S$) = 9.967\times10^{-6} \times $S$^{-3/2}$. The inverse of this rate gives a recurrence timescale as a function of fluence, i.e., $\tau($S$) = 1.003 \times10^5 \times $S$^{+3/2}$. The bolometric fluence of GRB\,221009A of 0.2\,erg\,cm$^{-2}$ has a recurrence rate at Earth of 9,700\,years. 

We can repeat this measurement while accounting for the uncertainty on the fluence of GRB\,221009A and on the bursts used in the fit, either directly or by sampling the scaling distributions in Appendix\,\ref{app:bolo_scalings}. Accounting for these uncertainties, a fit to the logN-logS distribution with a fixed -3/2 index gives a recurrence time of 9,200\,years and an uncertainty range of 7,200-11,200\,years. Uncertainties in this paragraph are reported for the 80\% confidence interval (i.e., 90\% lower and 90\% upper bounds). Allowing both scale and index to vary gives the measured index reported in Figure\,\ref{fig:fluence}, confirming we are in the regime where -3/2 power-law is valid, and a corresponding median recurrence rate of 7,400\,years and range 3,300-16,400\,years.

It is difficult to estimate this number by a combined logN-logP given the different peak interval timescales. We measure it, neglecting errors, by scaling the 1.024\,s peak flux of GRB\,221009A to the peak flux intervals of the individual instrument distributions according to the procedure described in Appendix\,\ref{app:peak_flux_scalings}. The values from Konus, GBM, and BATSE span 11,300-15,500 years, suggesting the peak flux recurrence rate is even more extreme. The individual instrument fluence recurrence rates (including \textit{PVO} and \textit{Vela}) span 6,100-9,600 years. No matter how the recurrence rate is measured, its brightness at Earth occurs on the order of one time every $\sim10,000$ years.

This value is substantially rarer than previous reports in the literature. The first reported rarity calculation in \citet{2022GCN.32793....1A} placed the recurrence rate at roughly every 500\,years, which helped motivate the original BOAT claims. \citet{Malesani2023redshift} and \citet{o2023structured} perform a calculation similar to our method with claimed recurrence rates of 22-122 and 300-1,100\,years, respectively. In these three cases our more stringent claim is largely driven by our use of the proper fluence value after accounting for instrumental effects, as the recurrence rates are a very strong function of brightness. \citet{williams2023grb} utilize the collapsar evolution model from \citet{lien2014probing} to place a lower limit of $\sim1,000$ years on the recurrence rate. Their use of a proper fluence value may suggest GRB\,221009A is unique compared to the broader modeled population, though we note their lower limit is consistent with our measure.

While GRB\,221009A is truly unique by observed brightness at Earth, it may not be unique by intrinsic brightness measures in the full observed prompt sample. There are more than 10,000 observed prompt GRBs, with only $\sim1,000$ well-localized ($\sim$few arcminutes), but fewer than 500 bursts have determined isotropic-equivalent energetics values. In these 500, there are $\sim20$ bursts with an $E_{\rm iso}$ within a factor of 5 of GRB\,221009A, and $\sim3$ bursts within a factor of 2. As the observed prompt sample is 20 times larger, it is likely that there are bursts in the observed sample with E$_{\rm iso}$ greater than GRB\,221009A, but whose redshift or broadband spectra were not measured. There are two known GRBs with higher L$_{\rm iso}$ than GRB\,221009A and we similarly expect more to remain unidentified in the broader sample. %There is a similarly small number of analogs in the observed sample (Section\,\ref{sec:analogs}) suggesting GRB\,221009A is on 

\subsection{Why GRB\,221009A is the BOAT}
GRB\,221009A is certainly the BOAT at Earth. This is easily explained as being an intrinsically bright burst in unusual proximity to Earth. This requires only a particularly rare event. GRB\,221009A was identified in a surprisingly small comoving volume, being 100 times smaller than the volume within which comparable bursts have been identified and 1000 times smaller than the volume it would have been detected within; this oddity is exacerbated when accounting for the declining source rates of collapsars since redshift $\sim3$ \citep{lien2014probing}. Further, it is extremely surprising that this burst is also the prompt E$_{\rm iso}$ record holder. One possible observational bias is that much of the emission of the burst would not be recovered at Earth if GRB\,221009A occurred at $z\approx1$ or greater. However, the main emission episode is detectable deep into the universe and alone contains enough energy to be the record E$_{\rm iso}$. 

The rarity of the event is not an explanation for why GRB\,221009A is the $E_{\rm iso}$ record holder. One part of the explanation is the unusually high bulk Lorentz factor in this burst, inferred either through pair opacity arguments or requiring the GBM emission to be optically thin \citep{grb221009a_lesage_fermi_gbm_2023}. In contrast to the prompt L${\rm iso}$ being near the record, the afterglow luminosity across the electromagnetic spectrum is within the observed distributions \citep{williams2023grb,2023arXiv230204388L,kann2023grb}. An unusually narrow jet opening angle could produce an intrinsically bright prompt signature while resulting in a more typical total kinetic energy in the jet. For GRB\,221009A to match the highest collimation-corrected total energetics in the Konus sample \citep{tsvetkova2017konus,tsvetkova2021konus} a collimation correction factor of $\sim1,000$ is required, corresponding to a top-hat jet half-opening angle constraint of $<2.6^\circ$. 

Making the strong assumption that the collimation-corrected prompt energetics of GRB\,221009A are not in excess of the known collapsar class distributions, these observations allow us to understand what could have occurred to produce this burst. Either due to unusual conditions in jet formation and propagation or in the progenitor star and circumburst properties, the jet core achieved particularly high velocity while remaining very tightly collimated. This collimation was maintained, despite a long-lived accretion phase, over 10 orders of magnitude in size from escape at the surface of the star to the external shock radius. This concentrated, highly energetic jet core was ideally aligned towards Earth.

Thus, either GRB\,221009A has a concentrated, particularly energetic jet core or GRB\,221009A must also hold the collimation-corrected prompt energetics record, which would be all the more exceptional. A reasonably modeled jet structure and opening angle will give temporal and spectral index closure relations that are well-matched by observations across the electromagnetic spectrum, providing a self-consistent check. The exceptional dataset for this burst must contend with Galactic extinction due to alignment with the plane of the Milky Way, host-galaxy extinction and contamination, and the expected supernova. Early results also utilize incomplete datasets. Below we place the prompt results in context with initial afterglow modeling of this burst, highlighting that a self-consistent picture has not emerged. Thus, a full understanding of why GRB\,221009A is the (prompt) BOAT is not yet possible.

\citet{williams2023grb} argue for a narrow jet using energetics arguments and report a steepening of the X-ray afterglow at $8\times10^4$\,s which would correspond to a half-jet opening angle of $\sim2^\circ$. However, the observed temporal decay in X-rays is inconsistent with expectations for a top-hat jet. \cite{2023arXiv230204388L} include radio data and model the burst as occurring in a wind medium with a narrow jet opening angle of $\sim1.5^\circ$. The corresponding kinetic energy E$_k \approx 4 \times 10^{50}$\,erg, which is within the known distribution. They note the inability to fully match typical closure relations, and this model struggles to explain the lack of a jet break seen in optical lightcurves. A top-hat jet with opening angle of $1.5^\circ$ would give a collimation-corrected E$_{\gamma}\approx4\times10^{51}$\,erg which is within the normal distributions \citep{tsvetkova2017konus,tsvetkova2021konus}. However, these E$_k$ and E$_{\gamma}$ values imply an unrealistic prompt gamma-ray efficiency of 90\%. 

A significantly narrower jet is disfavored with fiducial assumptions due to upper limits on the afterglow polarization \citep{negro2023ixpe}, though a narrower jet is possible if the typical assumed values for microphysical jet parameters are altered. \citet{kann2023grb} note no jet break in the optical data and infer a lower-limit on the jet opening angle of $\sim10^\circ$ \citep{kann2023grb}. Assuming a top-hat jet with this value gives $E_\gamma\approx2\times10^{53}$\,erg which would be substantially higher than any prior measured value \citep{cenko2011hyper,tsvetkova2017konus}. 

For previous particularly bright bursts, some have argued for a two-component jet model with a narrower ultrarelativistic jet surrounded by a wider jet with lower energy
\citep[e.g.][]{berger2003common,sheth2003millimeter,racusin2008broadband,kann2018optical}. Such a model may explain the nominally conflicting results reported and has been invoked for GRB\,221009A \citep{sato2022two}. \citet{o2023structured} invoke a multi-component jet model with an angular energy density dependence, declining as a broken power-law. This broadly explains the afterglow data, but would still require a record E$_K$ and E$_{\gamma}$.

The picture is further complicated by potential contamination by the expected supernova signal. \citet{grb221009a_fulton_optical_2023} assume the optical and X-ray emission have no synchrotron break between them and model the differing temporal decays as arising from the emergence of the supernova. \citet{shrestha2023lack} argue a supernova cannot explain the temporal decay differences between these wavelengths and argue for no bright supernova. \citet{grb221009a_jwst_hst_2023} find the JWST spectra at $\sim$two weeks after the prompt emission to be consistent with a power-law, again arguing against a bright supernova. They model the afterglow with an early jet break (and thus a narrow jet) and with a spectral break between optical and X-rays, which differs from other analyses.

\subsection{GRB\,221009A at Greater Distances}
For comparison of GRB\,221009A to other bright GRBs, it is useful to consider how it would appear at greater redshifts more typical of the observed collapsar sample. In Figure\,\ref{fig:eiso} a dashed line is overlaid providing an approximate trigger threshold for \textit{Fermi}-GBM, assuming that cosmological redshift effects on duration and observed energy are effectively counteracted by \enquote{tip-of-the-iceberg} effects from recovering only the bright, hard peaks of more distant bursts \citep{kocevski2013lack,moss2022instrumental}. It is visually evident that this assumption is reasonable. The GRB\,221009A prompt emission is comprised of a triggering pulse, the main emission, and the last bright pulse at $\approx$T0+500\,s. The initial pulse would trigger \textit{Fermi}-GBM to $z\approx1.3$ \citep{grb221009a_lesage_fermi_gbm_2023}, beyond which it would trigger at the onset of the main pulse. The main emission would still be recovered beyond the highest redshift measured for any long GRB.

\subsection{Implications for High Redshift GRBs}
High redshift GRBs, particularly those above $z\sim6$, could be used to study early evolution of galaxies, to probe reionization, and to study metallicity from the death of the first stars \citep{tanvir2021exploration}. GRB\,221009A could have been detected well beyond $z\sim10$. At these distances the observed trigger would be on the pulse leading up to the brightest intervals, with a peak energy of $\sim1-3$\,MeV \citep{grb221009a_lesage_fermi_gbm_2023,grb221009a_konus_2023}. At a redshift of 10 this would be observed with an E$_{peak}$ of $100-300$\,keV. Other GRBs detected beyond $z\approx6$ show similar observed spectral hardness, e.g., the recent GRB\,210905A at $z\sim6.3$ and peak energy of 145\,keV \citep{2022A&A...665A.125R}. Thus, a population of GRBs with peak observed energies at Earth in the hundreds of keV should exist, which should be accounted for in the design of high-$z$ GRB missions \citep{white2021gamow,amati2021theseus}. 

\subsection{Comparison with Ultra-long GRBs}
Ultra-long GRBs may be the longest duration events belonging to the extreme tail of the long GRB sample, or they may be a distinct class with longer-lived central engines than typical collapsars. The threshold for inclusion in the ultra-long class is not agreed upon, we here explore thresholds of 1000\,s and 3,600\,s. Beyond $z\approx0.7$ GRB\,221009A would generally have an inferred duration beyond 1000\,s owing to cosmological time dilation. For GRB\,221009A to have a measured burst duration of longer than 3,600\,s \citep{kann2018optical} it would need to be beyond $z\approx8.6$. For collapsars with measured redshift $\sim75$\% are beyond $z\approx0.7$ while few are beyond $z\approx8.6$. GRB\,221009A may or may not be a member of the putative ultra-long GRB sample, depending on the threshold value assumed. 

The Konus-\textit{Wind} ultra-long GRB sample (D. Svinkin, in prep.) is the most complete of any instrument, with nearly two dozen events beyond a 1,000\,s threshold. The highest fluence values of these bursts are $\sim5-6\times$10$^{-4}$\,erg\,cm$^{-2}$ for GRB\,080407 and the record duration burst GRB\,111209A. Thus, none of the Konus-\textit{Wind} ultra-long GRBs are remotely as bright as GRB\,221009A. The highest peak flux values reach only $\sim1\times10^{-5}$\,erg\,s$^{-1}$\,cm$^{-2}$ for GRB\,080407 and GRB\,961029. The typical ratio of peak flux to fluence for Konus ultra-long GRBs is $\sim3$\% while GRB\,221009A crosses the 10\% boundary. This may suggest GRB\,221009A as intermediate between typical long and ultra-long GRBs, providing some support for a single continuum. There are additional GRBs that may fall into a similar range, including GRB\,840304 (discussed next) and GRB\,210905A with a duration of 870\,s at a redshift of $z=6.3$ \citep{2022A&A...665A.125R}.

\subsection{Analogs}\label{sec:analogs}
With the deep search of prompt GRB detections a few analogs to GRB\,221009A have been identified. A close analog is GRB\,990123. This burst is in the top 15 highest E$_{\rm iso}$ (Table\,\ref{tab:eiso}), the L$_{\rm iso}$ is $3-5\times10^{53}$\,erg/s, the kinetic energy ($1-5\times10^{50}$\,erg) is comparable to the narrow-jet one for GRB\,221009A \citep{2023arXiv230204388L}, and the half-jet opening angle is an unusually narrow $\sim$2$^\circ$ \citep{2006ApJ...637..889Z,laskar2013grb,tsvetkova2017konus}. The burst occurs at $z=1.604$ \citep{kelson1999iau,1999GCN...219....1H}, where the precursor pulse for GRB\,221009A would not be recovered by GBM. More speculatively, the light curve resembles the onset of the main emission episode of GRB\,221009A \citep{briggs1999observations}.

GRB 160625B is perhaps the strongest analog when considering the full lightcurve: both have a weaker triggering pulse, quiescence for $\sim$175\,s where the main emission occurs, with additional variable emission at $\sim600$\,s. GRB\,160625B occurred at a redshift of $z=1.406$ \citep{2016GCN.19600....1X} and has a comparable L$_{\rm iso}$. Both triggering pulses have particularly soft indices but the spectral curvature occurs more than an order of magnitude lower in GRB\,160625B compared to GRB\,221009A \citep{zhang2018transition,grb221009a_konus_2023, grb221009a_lesage_fermi_gbm_2023}, while the GRB\,160625B pulse has a far higher luminosity. %Further comparison of these bursts may be particularly interesting.

The only ultra-long analog is GRB\,840304 which is a top 5 burst by fluence, average burst by peak flux, and more than 1,000\,s long with two bright pulses followed by smooth emission \citep{klebesadel1984unusual}. This profile sounds similar to GRB\,221009A and may include afterglow in the duration calculation. Work is on-going to find this data.

\section{Conclusion}\label{sec:conclusion}
Prompt observations of GRB\,221009A allow for a number of advancements in our understanding of these extreme events. We have here studied what can be learned from a comparison against the full detected prompt GRB sample, including strengthening some results already understood in \citet{grb221009a_konus_2023} and \citet{grb221009a_lesage_fermi_gbm_2023}.

Our results are summarized as
\begin{itemize}
\item GRB\,221009A is the BOAT by three of the four measures of brightness. It is certainly the highest fluence and peak flux GRB ever identified, by a large margin. It is the highest E$_{\rm iso}$ burst ever identified and at the 99th percentile of L$_{\rm iso}$. These intrinsic extremes cannot be explained by observational bias.
\item While we have not directly measured higher E$_{\rm iso}$ in a GRB, it is likely that some of the prompt GRB detections, without known redshift or broadband spectra, have total energetics that exceed this burst.
\item We additionally explored the observation of this GRB had it occurred at greater distances, with implications for both ultra-long GRBs and high redshift GRBs. 
\item We have here explored why GRB\,221009A is so bright in the prompt emission, contributing to the advancement of understanding of this event. We identify three potential analogs whose joint study may prove fruitful.
\item We are unlikely to observe another event of such extreme brightness at Earth given the recurrence time on the scale of 10,000\,years. There is a reasonable chance this is the brightest burst at Earth since civilization began. If this rate calculation is correct, there is likely no brighter GRB signal within thousands of light years. At any given time, only a few dozen such plane waves of intense radiation of similar or even higher intensity are traversing though the Milky Way.
\end{itemize}

\section*{Acknowledgements}
We acknowledge the Universe for timing this burst to arrive at Earth after the invention of GRB monitors but during our active research careers. Our token optical astronomer would like to complain about the alignment with the Galactic plane, and requests the next one avoid this issue. The paper is dedicated to all the unsung publications that make population analyses like this work possible, particularly those which we missed.  

This paper is additionally dedicated to D. Alexander Kann who unexpectedly passed away during the review of this paper. The above acknowledgements are left unchanged as they were his suggestions. His expertise and input were key to this paper, and many others. It is a small consolation to add this dedication to a paper which may be read in several thousand years, when the fluence record is broken. Alex will be sorely missed.

We thank Eve Chase and Chris Fryer for putting some key authors in contact, allowing for the largely complete dataset. We thank the referee for prompt and valuable input.

J.~F.~Ag\"u\'i~Fern\'andez acknowledges support from the Spanish Ministerio de Ciencia, Innovaci\'on y Universidades through the grant PRE2018-086507.
D.~A.~Kann acknowledges the support by the State of Hessen within the Research Cluster ELEMENTS (Project ID 500/10.006). R.~Hamburg acknowledges funding from the European Union's Horizon 2020 research and innovation program under the Marie Skodowska-Curie grant agreement No. 945298-ParisRegionFP.

\bibliography{references}

\appendix

\section{Fluence and Peak Flux Bolometric Scalings}\label{app:bolo_scalings}
A major analysis portion of this paper is the conversion of reported fluence and peak flux values for GRB monitors which operated and reported in energy ranges narrower than the bolometric range of interest here. Konus values are already sufficiently close to bolometric. For BATSE and GBM bursts where the best-fit spectral form constrains curvature we directly integrate the fit parameters over the bolometric energy range. For \textit{Vela}, \textit{PVO}, and BATSE and GBM bursts best-fit by a power-law (where extrapolation would overestimate the true bolometric values), we must apply a scaling distribution to convert from one energy range to another. These scaling distributions can be constructed by taking the scaling values for a large sample of GRBs with measured spectral curvature from one energy range to another. For this we use the GBM 10 Year Spectral Catalog \citep{poolakkil2021fermi} which contains a complete spectral analysis to determine the best-fit spectrum over the full T$_{90}$ interval and the peak-flux interval, set to the 1.024\,s timescale for long GRBs. 

\begin{figure*}[!ht]
\centering
	\includegraphics[width=1.0\textwidth]{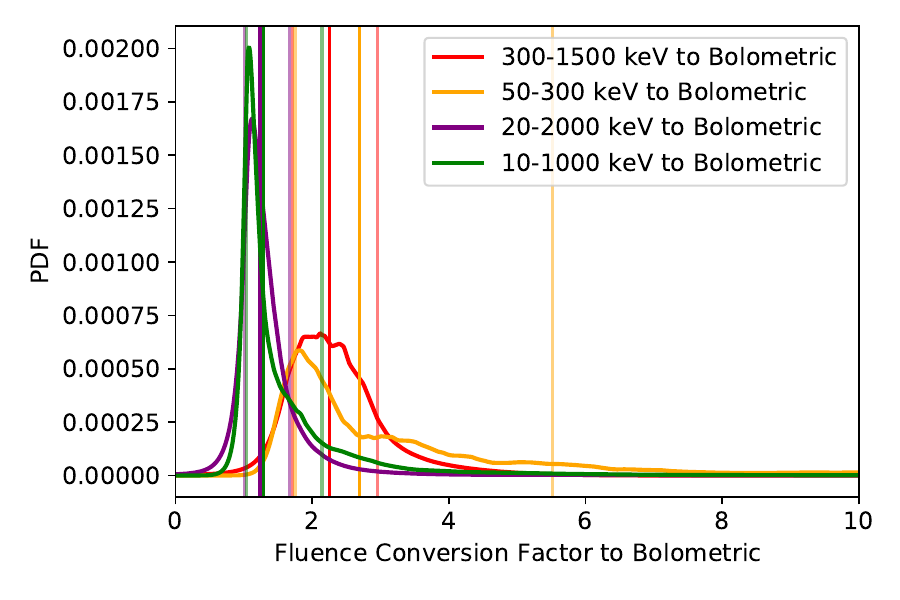}
    \caption{The PDF of scaling fluence as measured in a given energy range to the 1\,keV-10\,MeV bolometric energy range.
    }
    \label{fig:scale_bolo_fluence}
\end{figure*}

For the fluence sample there are 1,938 bursts considered. 376 are best-fit by a power-law; 1,562 are best-fit by a model with constrained curvature, i.e., a Comptonized function, BAND, or a smoothly-broken power-law.  The GBM fit is performed over the $\sim8$\,keV-39\,MeV energy range, meaning this is a slight extrapolation on the low end but otherwise within the GBM bandpass. For these 1,562 bursts we determine the scaling factor and uncertainty to convert between energy ranges on a burst-by-burst basis. These measures are constructed into a distribution, which gives the averaged scaling and uncertainties. For conversion of the GBM 10-1000\,keV fluence to the bolometric band we get a scaling value of 1.28$_{-0.24}^{+1.11}$, with the full distribution shown in Figure\,\ref{fig:scale_bolo_fluence}.

We can repeat the same procedure except instead of converting the 10-1000\,keV fluence to the bolometric, we convert it to the reported energy ranges for the other instruments, i.e., 300-1,500\,keV for \textit{Vela}, 50-300\,keV for \textit{PVO} (utilizing the values reported in \citealt{pvo_fenimore_2023}), and 20-2,000\,keV for BATSE. For the \textit{Vela} comparison we remove bursts with E$_{peak}<$300\,keV, necessary to avoid over-correction. The final scaling factors are the convolution of two of these distributions, i.e., for \textit{Vela} the final values convolve the 10-1,000\,keV to bolometric energy range with the inverse of the 10-1,000\,keV to 300-1,500\,keV energy range, allowing for a mapping from the 300-1,500\,keV fluence to the bolometric range.

\begin{figure*}[!ht]
\centering
	\includegraphics[width=1.0\textwidth]{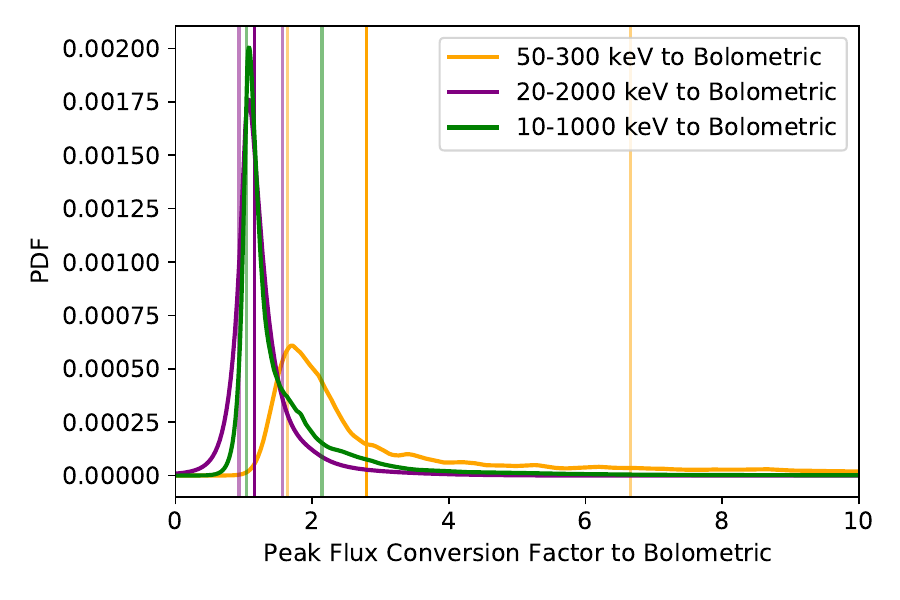}
    \caption{The PDF of scaling peak flux as measured in a given energy range to the 1\,keV-10\,MeV bolometric energy range. Note the particularly significant skew in the \textit{PVO} extrapolation (50-300\,keV), which may explain our anomalous \textit{PVO} peak flux values.
    }
    \label{fig:scale_bolo_flux}
\end{figure*}

For the peak flux intervals, the same procedure can be applied, with the results shown in Figure\,\ref{fig:scale_bolo_flux}. The peak flux scaling distributions are calculated based on the peak flux spectral fit parameter values in the GBM Spectral Catalog \citep{poolakkil2021fermi}. This is necessary given the generally harder spectra during peak flux intervals, as compared to the time-integrated fits, resulting in a slightly larger overall scaling value.

\section{Peak Flux Interval Scalings}\label{app:peak_flux_scalings}
In order to compare peak flux values between instruments to identify the brightest individual bursts we utilize parameters from the \textit{Fermi}-GBM Ten Year Catalog \citep{von2020fourth}. The GBM calculation of duration is the $T_{90}$ which is the time between the integrated 5\% and 95\% of the total burst fluence. This is determined by fitting time-resolved slices from pre-burst background to post-burst background, assuming a Comptonized function. An output of this analysis is a measure of the peak flux over 0.064\,s, 0.256\,s, and 1.024\,s intervals. 

\begin{figure*}[!ht]
	\centering
	\includegraphics[width=1.0\textwidth]{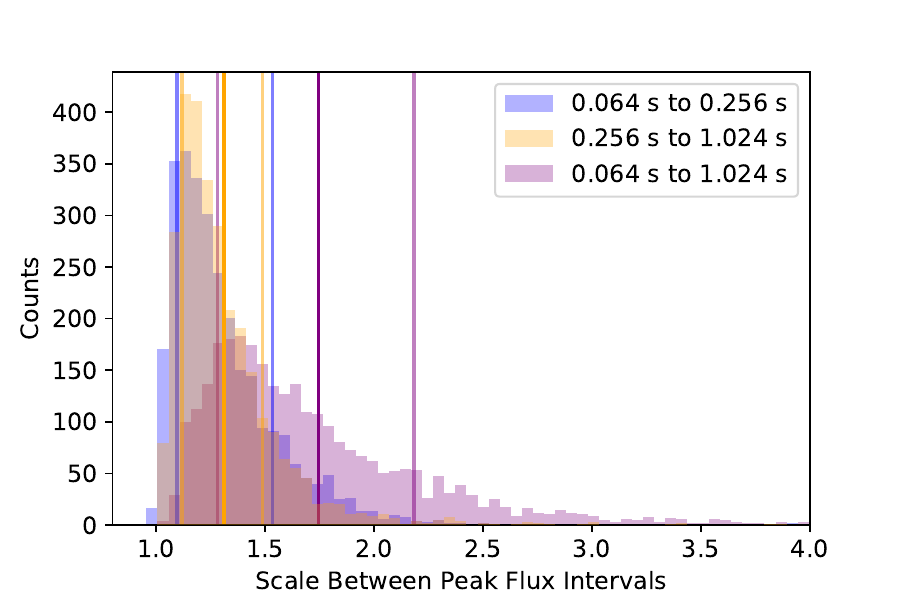}
    \caption{The distribution of the ratios of a given GBM peak flux interval against another. Vertical lines denote the median and $1\sigma$ bounds for each distribution. The 0.064\,s to 0.256\,s and 0.256\,s to 1.024\,s means are nearly identical and overlaid on the figure.
    }
    \label{fig:peak_flux_scalings}
\end{figure*}

From comparing the sample the 0.064\,s peak flux is $1.31_{-0.22}^{+0.22}$ times the 0.256\,s interval. The 0.256\,s peak flux value is $1.31_{-0.19}^{+0.18}$ the 1.024\,s interval. Lastly, the 0.064\,s is $1.75_{-0.47}^{+0.44}$ the 1.024\,s values. These are mean values with $1\sigma$ uncertainties. The full distributions are shown in Figure\,\ref{fig:peak_flux_scalings}. Noting that $1.31^2=1.72$, we see that each factor of two in peak flux interval scales as $\sqrt{1.31}\approx1.15$, or $\sim15$\%. The scale invariance and reasonable error bars allow us to scale peak flux intervals from different instruments into a standardized range. While not valid on each individual burst, the population scalings are accurate. Individual variation will not affect our conclusions here as no burst is close to GRB\,221009A in peak flux and the L$_{\rm iso}$ measure is already bounded.

\end{document}